\documentclass{emulateapj}
\usepackage{amsmath}

\newcommand{\0}{\phantom{$0$}}
\newcommand{\p}{\phantom{$+$}}
\newcommand{\m}{\phantom{$-$}}
\newcommand{\h}{\phantom{'}}

\shorttitle{The Lens Redshift and Group Environment for HE~0435$-$1223}
\shortauthors{Morgan et al.}

\begin{document}

\title{The Lens Redshift and Group Environment for 
HE~0435-1223\altaffilmark{1}}
\author{N.~D.~Morgan\altaffilmark{2},
	C.~S.~Kochanek\altaffilmark{2},
	O.~Pevunova\altaffilmark{3},
	P.~L.~Schechter\altaffilmark{4}
}

\altaffiltext{1}{Based on observations obtained with the Magellan 
Consortium's Clay Telescope and the NASA/ESA {\it Hubble Space Telescope}.
{\it HST} observations are obtained at the Space Telescope Science Institute, 
which is operated by the Association of Universities for Research in 
Astronomy, Inc., under NASA contract NAS 5-26555.  These observations are 
associated with {\it HST} program 9744.}

\altaffiltext{2}{Department of Astronomy, Ohio State University, Columbus, OH 
43204; nmorgan@astronomy.ohio-state.edu, ckochanek@astronomy.ohio-state.edu}

\altaffiltext{3}{Infrared Processing and Analysis Center, M/S 100-22, 
California Institute of Technology, Jet Propulsion Laboratory, Pasadena, CA 
91125; olga@ipac.caltech.edu}

\altaffiltext{4}{Department of Physics, Massachusetts Institute of
 Technology, Cambridge, MA 02139; schech@space.mit.edu}

\begin{abstract}
The redshift of the galaxy lensing HE~0435$-$1223 is 0.4546 $\pm$ 0.0002, 
based on observations obtained with the Low Dispersion Survey Spectrograph 2 
(LDSS2) on the Magellan Consortium's 6.5~m Clay telescope.  {\it Hubble Space 
Telescope}/ACS observations of the system also reveal a spiral-rich group 
of 10 galaxies within 40\arcsec\ of the elliptical lensing galaxy.  The 
redshifts for two of these galaxies were measured to be in the foreground 
(at $z=0.419$) with respect to the lens, thus at least some of the 
nearby galaxies are not part of the same physical group as the lensing galaxy.
Mass models of the system (assuming same-plane deflectors) that take the local
group environment into account do better at explaining the observed 
emission-line flux ratios (which are presumably unaffected by microlensing) 
than single halo models, but the match is still not perfect.  In particular, 
component A (a minimum of the light travel time) is observed to be 0.20 
mag brighter than predicted and component C (also a minimum image) is observed
to be 0.16 mag fainter than predicted.  Mass models for the system predict 
an A--D time delay of either 15.8 or 17.6 days ($H_{o} = 72$ km s$^{-1}$ 
Mpc$^{-1}$) depending on the details of the local galaxy environment.
\end{abstract}

\keywords{gravitational lensing: individual (\objectname{HE~0435$-$1223})}

\section{Introduction}

For a variety of reasons, follow-up observations of gravitationally
lensed quasars have not kept pace with the lens discovery rate.
For example, both lens and source redshifts are required to convert a 
measured time-delay into an estimate of the Hubble constant (Refsdal 1964) or 
to measure the mass-to-light evolution of lensing galaxies (Rusin et 
al. 2003), but only half of the $\sim$80 known systems have complete redshift 
information.  Poor knowledge of the lens galaxy environment
also hinders lensing applications.  Nearby groups or clusters
can bias estimates of the Hubble constant by contributing
some fraction of the lensing mass convergence (Gorenstein, Shapiro \& Falco 
1988; Saha 2000), and such structures are likely responsible for the large 
($\sim10\%$) shears required to model quadruple systems (Holder \& Schechter 
2003) and for explaining the observed quad-to-double ratio (Keeton \& 
Zabludoff 2004).  It is therefore important to understand the characteristics 
of each lens system as thoroughly as possible.  In this paper, we report our 
measurement of the lens redshift for the quadruple quasar HE~0435$-$1223 and 
explore the system's local galaxy environment from {\it HST}/ACS imaging.

\begin{figure*}[t]
\plotone{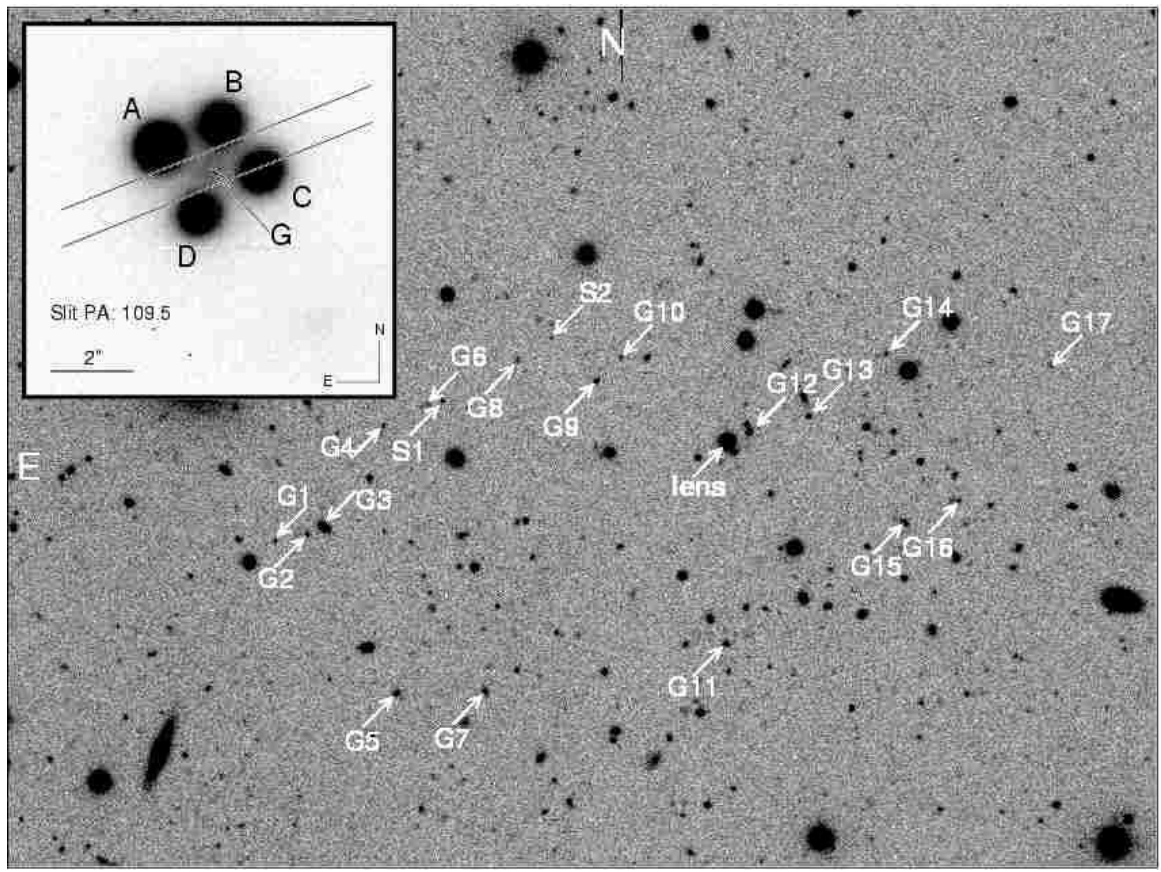}
\caption{{\it R}-band image of HE0435$-$1223 and field
obtained with the du Pont 2.5-m telescope at Las Campanas. The panel spans
7.1\arcmin\ across.  Targets observed with LDSS-2 are labeled (S1, S2, lens,
G1-17).  {\it Inset}: Close-up of HE~0435$-$1223 from Figure~2 of Wisotzki et
al. (2002).  The parallel lines show the slit orientation and width used for
LDSS2 spectroscopy.}
\label{fig:fc_dupont}
\end{figure*}
 
HE~0435$-$1223 was discovered to be a $z~=~1.689$ quasar during the 
Hamburg/ESO (HES) survey for bright quasars in the Southern Hemisphere 
(Wisotzki~et~al.~2000).  The quasar was later found to be gravitationally 
lensed by Wisotzki et al. (2002) during snapshot 
followups of HES quasars with the 6.5~m Baade telescope at the Las 
Campanas Observatory (LCO).  The image morphology is a symmetric quad 
with image separations of 2\farcs3 and 2\farcs6 along 
the long axes of the configuration, somewhat reminiscent of the 
Northern Hemisphere ``Einstein Cross'' Q~2237+0305 
(Huchra~et~al.~1985).  The lensing galaxy was prominently detected in the 
Baade discovery images, with its light profile and $gri$ aperture 
colors suggestive of a $0.3<z<0.5$ elliptical galaxy.  

The system holds high promise for measuring a differential time delay 
and estimating the Hubble constant.  Wisotzki~et~al.~(2002) cite evidence 
for both short-term (20\% over $\sim$2 months) and long-term (1 magnitude 
over $\sim$ 12 years) variability in the total quasar flux.  The symmetric 
image configuration also implies relatively short differential time delays 
(on the order of several weeks), which ought to make it straightforward to 
distinguish the time-delay signature from longer-timescale microlensing 
variability.

Wisotzki et al. (2003) provided a refined galaxy redshift estimate of 
$z=0.44 \pm 0.02$ based on integral field observations of the system.  
However, their estimate was obtained from low-resolution ($\Delta\lambda 
= 300$\AA) spectral rebinning that traced the galaxy's overall spectral 
energy distribution rather than from the detection of stellar absorption 
features.   In \S\ref{ldss2}, we report a precise spectroscopic measurement 
of the lensing galaxy redshift of $z=0.4546\pm0.0002$ using the 
Low-Dispersion Survey Spectrograph 2 (LDSS2) on the Clay 6.5~m telescope.  
In \S\ref{hstacs}, we make use recent {\it HST}/ACS observations of the 
system to explore the HE~0435$-$1223 galaxy environment and find evidence for 
a spiral-rich group of at least 10 galaxies within 40\arcsec, at least 
two of which are in the foreground (at $z=0.419$) with respect to the lens.  
Overall the data 
suggest a complex lensing environment.  We explore several lens models 
for the system using the {\it HST}/ACS astrometry and emission-line flux 
ratios from Wisotzki et al. (2003) in \S\ref{models}, and discuss the model 
implications and time-delay predictions in \S\ref{discuss}.

\section{Magellan/LDSS2 Observations}
\label{ldss2}

HE~0435$-$1223 was observed on 2002 December 11 using the Low Dispersion 
Survey Spectrograph 2 (LDSS2; Allington-Smith et al. 1994) at the Magellan 
Consortium's Clay telescope at LCO.  The LDSS2 is a multi-slit spectrograph 
with a 7\arcmin\ diameter field of view and 0\farcs378 pixel$^{-1}$ CCD 
detector.  We used the medium-red grism blazed at 6000~\AA, which provided a 
nominal dispersion of 5.3~\AA\ pixel$^{-1}$ and a useful wavelength coverage 
of 4500~\AA\ to 8500~\AA.  A slit width of 0\farcs74 was used for all
targets when constructing the aperture mask.  The night was photometric and 
the Shack-Hartmann wavefront sensor helped to deliver an image quality of 
$\sim$0\farcs5 FWHM.

In addition to the lensing galaxy, we obtained simultaneous spectra of 19
objects within 3\arcmin\ of HE~0435$-$1223 (see Figure~1).  Our observing 
sequence consisted of two 30 minute exposures with slits
centered on the lensing galaxy and surrounding field galaxies, followed
by two 10 minute exposures through the same mask but offset
by 1\farcs25 to the North-East.  These last two observations
were used to obtain quasar template spectra of components A and B 
(Figure~1 inset), which were subsequently used to remove quasar spillover 
from the lens galaxy spectrum.  No spectrophotometric standards were 
observed.

\begin{deluxetable*}{cllllcc}
\tablecaption{Redshift Analysis for HE~0435-1223 Lens and Field Galaxies
\label{TABLE1}}
\tablewidth{0pt}
\tablehead{  
\colhead {Object}&
\colhead {R.A.}&
\colhead {Dec.}&
\colhead {z}&
\colhead {Line Identification}&
}
\startdata
  G1\dotfill & 4 38           26.03 & $-$12 18 \phantom{0}4.6 & 0.8124 $\pm$ 0.0001 & [\ion{O}{2}]/H$\beta$/[\ion{O}{3}]  \nl 
  G2\dotfill & 4 38           25.25 & $-$12 18 \phantom{0}1.2 & 0.3183 $\pm$ 0.0003 & [\ion{O}{2}]/H$\beta$/[\ion{O}{3}]  \nl 
  G3\dotfill & 4 38           24.85 & $-$12 17           58.1 & 0.3021 $\pm$ 0.0002 & [\ion{O}{2}]/[\ion{O}{3}]/H$\alpha$ \nl 
  G4\dotfill & 4 38           23.54 & $-$12 17           19.0 & 0.3915 $\pm$ 0.0005 & [\ion{O}{2}]/H$\beta$/[\ion{O}{3}]  \nl 
  G5\dotfill & 4 38           22.69 & $-$12 18           57.4 & 0.6240 $\pm$ 0.0007 & [\ion{O}{2}]/[\ion{O}{3}]           \nl 
  G6\dotfill & 4 38           22.48 & $-$12 17 \phantom{0}9.9 & 0.6676 $\pm$ 0.0004 & [\ion{O}{2}]/[\ion{O}{3}]           \nl 
  G7\dotfill & 4 38           20.46 & $-$12 18           54.1 & 0.5580 $\pm$ 0.0001 & [\ion{O}{2}]/H$\beta$               \nl 
  G8\dotfill & 4 38           20.32 & $-$12 16           50.7 & 0.1841 $\pm$ 0.0002 & H$\beta$/[\ion{O}{3}]               \nl 
  G9\dotfill & 4 38           18.28 & $-$12 16           56.2 & 0.3380              & [\ion{O}{2}]?                       \nl 
 G10\dotfill & 4 38           17.73 & $-$12 16           46.9 & 0.3691              & [\ion{O}{2}]?                       \nl 
 G11\dotfill & 4 38           14.50 & $-$12 18           29.2 & 0.4579 $\pm$ 0.0001 & [\ion{O}{2}]/H$\beta$/[\ion{O}{3}]  \nl 
 G12\dotfill & 4 38           24.85 & $-$12 17           58.1 & 0.4191 $\pm$ 0.0002 & [\ion{O}{2}]/H$\beta$/[\ion{O}{3}]  \nl 
 G13\dotfill & 4 38           12.88 & $-$12 17 \phantom{0}2.6 & 0.4189              & [\ion{O}{2}]?                       \nl 
 G14\dotfill & 4 38           11.06 & $-$12 16           37.7 & 0.4872 $\pm$ 0.0005 & [\ion{O}{2}]/CaII H\&K              \nl 
 G15\dotfill & 4 38           10.26 & $-$12 17           39.4 & 0.3766 $\pm$ 0.0001 & [\ion{O}{2}]/[\ion{O}{3}]           \nl 
 G16\dotfill & 4 38 \phantom{0}8.93 & $-$12 17           29.9 & 0.6880 $\pm$ 0.0001 & CaII H\&K                           \nl 
 G17\dotfill & 4 38 \phantom{0}6.90 & $-$12 16           37.5 & 0.8434              & [\ion{O}{2}]?                       \nl 
     & & & & \nl
lens\dotfill & 4 38           14.87 & $-$12 17           14.8 & 0.4546 $\pm$ 0.0002 & CaII H\&K                           \nl 
     & & & & \nl
  S1\dotfill & 4 38 22.13 & $-$12 17 \phantom{0}8.3 & 0 & Star \nl 
  S2\dotfill & 4 38 19.52 & $-$12 16 41.6 & 0 & Star \nl 
\enddata
\tablecomments{Redshifts for galaxies observed with LDSS2.  IDs correspond to labels in Figure~1.  
Redshift errors are the rms dispersion among multiple lines (when available), and do not include the 
$\delta z=0.0002$ uncertainty in the underlying wavelength calibration.  All positions are J2000.0
coordinates.}
\end{deluxetable*}

\begin{figure}[b]
\plotone{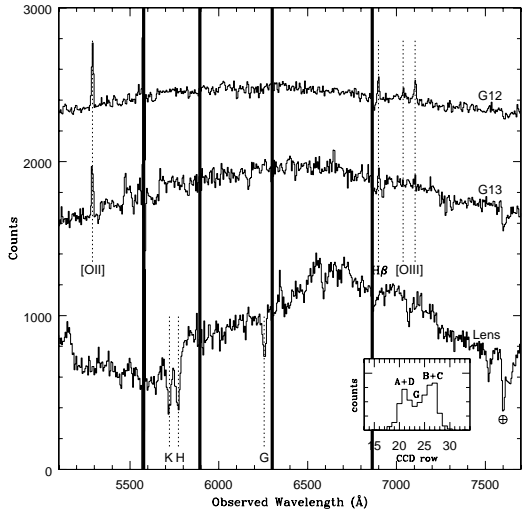}
\caption{Spectra of the lensing galaxy (bottom) and neighboring galaxies G12
an G13 (top and middle).  The lens spectrum is shown after subtracting 0.2
times component A's spectrum.  The four vertical strips denote skylines
used for wavelength calibration.  The lens, G13, and G12 spectra have been
shifted by 100, 1500, and 2250 counts, respectively, for clarity.
{\it Inset:} Slice along the spatial direction for the lens galaxy spectrum.
The two peaks correspond to spillover flux from the A+D and B+C quasar images,
with the lens galaxy in the middle.}
\label{fig:spectra}
\end{figure}

\begin{figure}[b]
\plotone{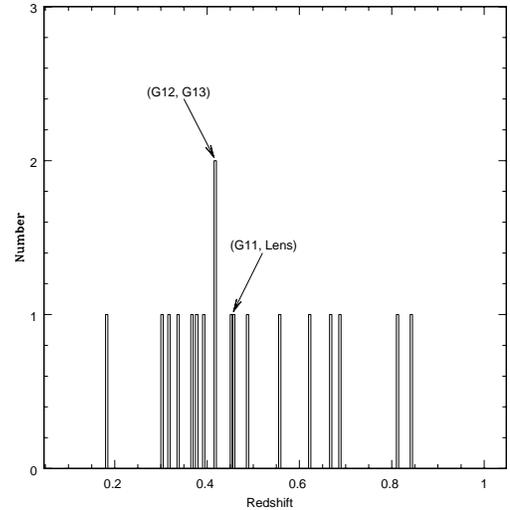}
\caption{Redshift histogram of galaxies observed with the LDSS2.\ \\ \ \\ \ \\ \
 \\ \ \\ \ \\ \ \\}
\label{fig:hist}
\end{figure}

Initial data reduction consisted of bias-subtraction, flat-fielding,
and removing cosmic-rays by interpolating from neighboring pixels.  The 
two-dimensional spectra were then averaged together to produce a single
on-galaxy and off-galaxy image.

\begin{figure*}[t]
\plotone{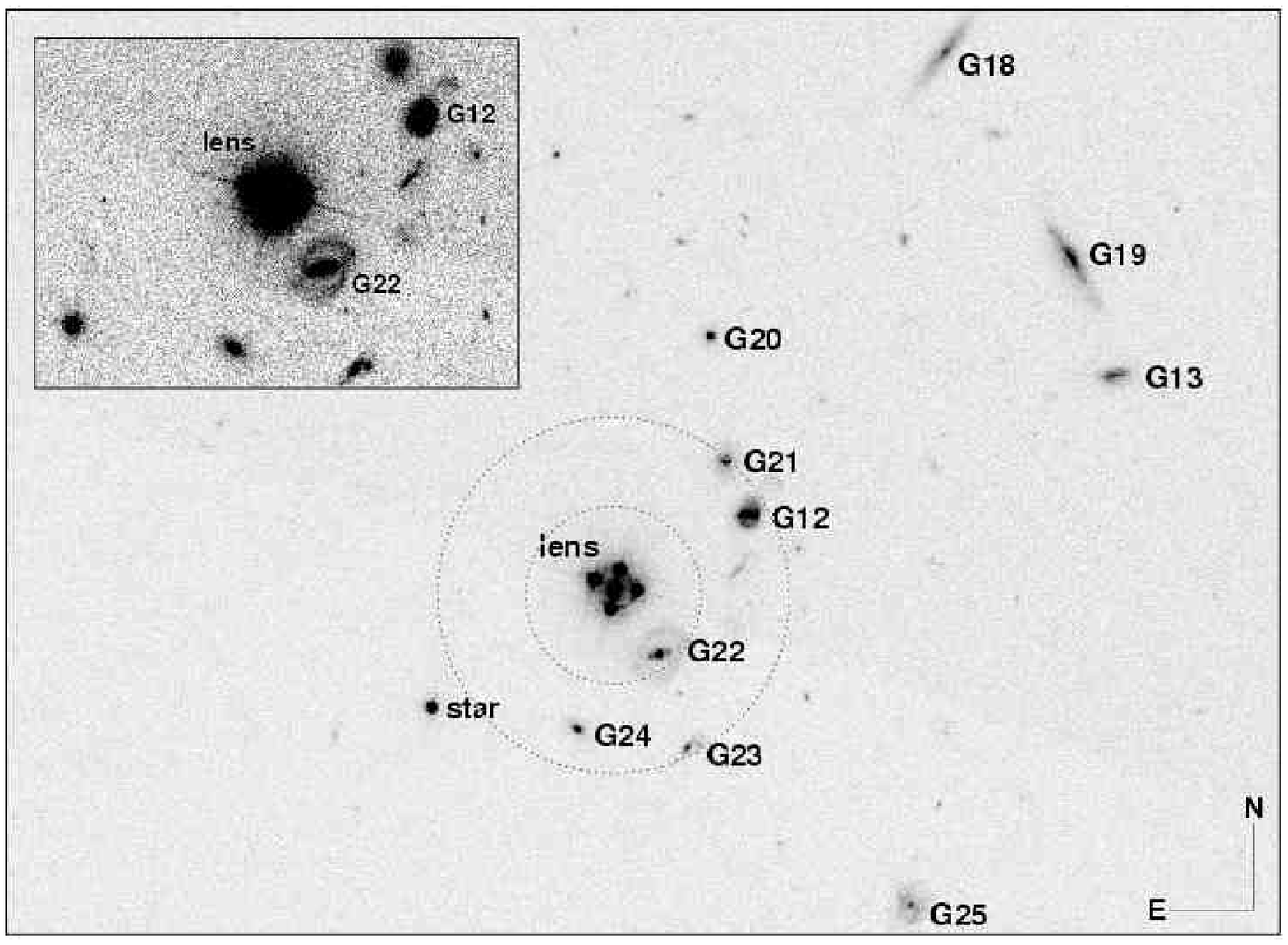}
\caption{PyDrizzled {\it HST}/ACS F814W image of HE~0435$-$1223 and field.
Prominent nearby galaxies are labeled.  The concentric circles centered
on the lensing galaxy mark 5\arcsec\ and 10\arcsec\ radii.  Inset shows
the inner 10\arcsec\ region (centered on the lens) at high contrast.}
\label{fig:fc_hst}
\end{figure*}

Since these are two-dimensional spectra, care must be taken when deriving
a wavelength solution and performing the extraction.  The wavelength
calibration was obtained using night sky lines bracketing each target spectrum.
We identified eight lines from the spectral atlas of Osterbrock~et~al. (1996),
providing roughly uniform coverage between [\ion{O}{1}]$\lambda$5577 and 
OH$\lambda$8827.  The lines were identified on 2-4 dispersion rows (depending 
on the target slit length) with two rows always straddling the target spectrum,
providing 16-32 points of known wavelength as a function of CCD row and 
column.  We then fit a fourth order polynomial along the dispersion direction 
and a linear fit along the spatial direction, which provided the wavelength 
solution as a function of pixel position on the chip.  The wavelength solution
with respect to the reference lines was extremely good: rms better than 0.8 
\AA\ (0.15 pixels) for all spectra.  The sky background was then removed by 
fitting and subtracting a line to the sky level for each CCD column.

\begin{figure*}[t]
\plotone{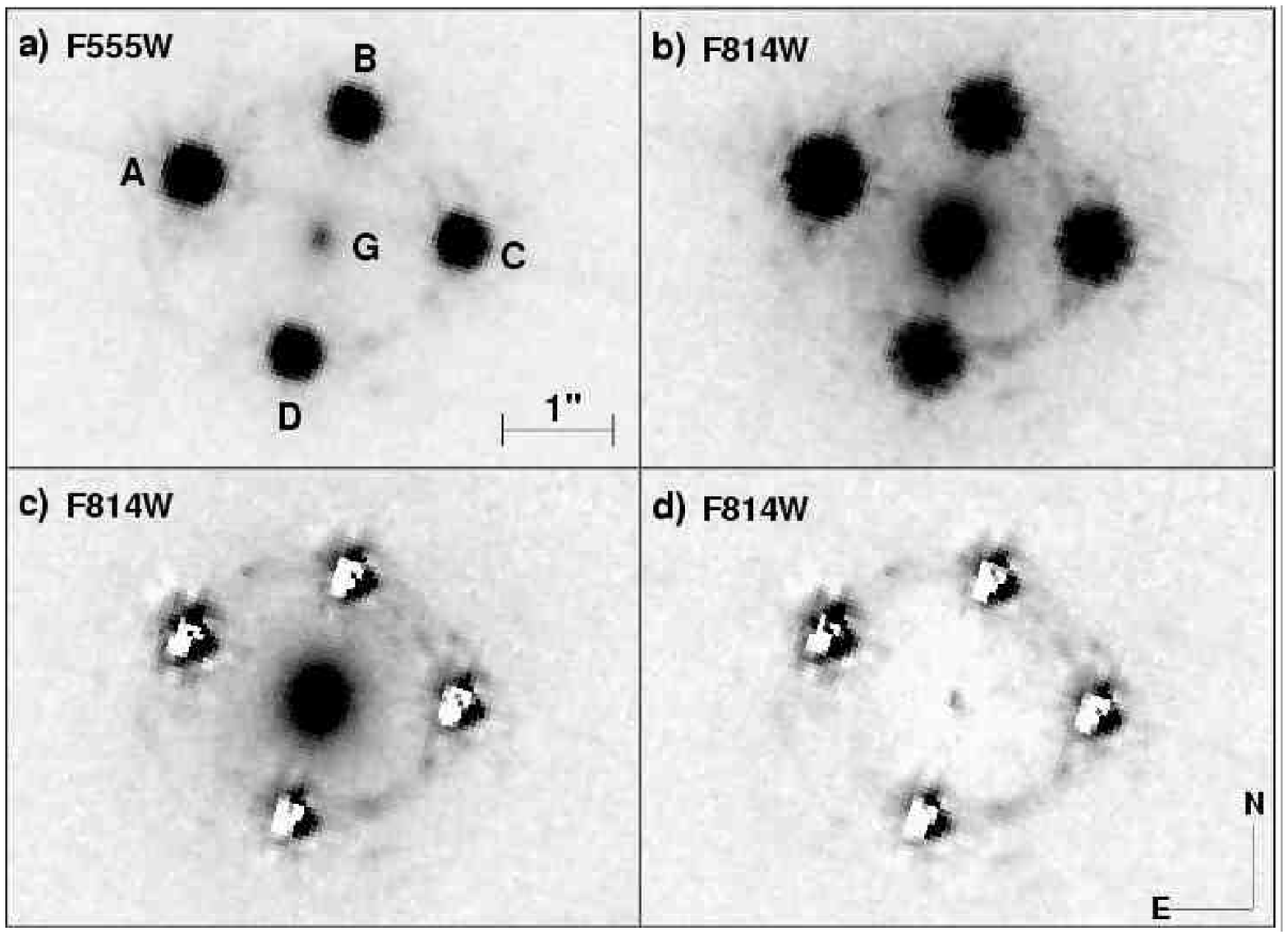}
\caption{({\it a}): PyDrizzled {\it HST}/F555W observations of HE~0435$-$1223.
({\it b}): PyDrizzled {\it HST}/F814W observations of HE~0435$-$1223.
({\it c}):~Same as ({\it b}), but after subtracting the best four-component
PSF model.  ({\it d}): Same as ({\it b}), but after subtracting the best
four-component PSF plus de Vaucouleurs model for the lensing galaxy.
Contrast in all panels is from --0.02 $e^{-}s^{-1}$ to 0.5 $e^{-}s^{-1}$.}
\label{fig:hstvi}
\end{figure*}

The spectral dispersion was mostly parallel to the CCD rows, but did bend
by several pixels in the vertical from blue to red.  To define the extraction 
path for the lensing galaxy, we fit two overlapping Gaussian profiles to each 
CCD column to model the flux from the A+B spectra.  A second-order polynomial 
fit to A's center as a function of chip column gave a good fit to 
the dispersion drift (fit rms of 0.08 pixels), and was subsequently used as 
the extraction path for both the quasar and lensing galaxy spectra.  The 
extraction was performed using a three pixel (1\farcs1) width for quasar 
components A and B, and a two pixel (0\farcs76) width for the lensing galaxy.
For the remaining targets, we used a wide extraction width of 6 pixels 
(2\farcs3) parallel to the CCD rows.

Figure~2 shows the extracted lens galaxy spectrum after subtracting 0.2
times component A's spectrum, with the scale factor chosen such that the
blueward side of the \ion{Mg}{2} broad emission line at $\sim$7500 \AA\ 
appeared smoothly subtracted by eye.  The \ion{Ca}{2} H\&K absorption features 
are visible, as well as the 4000 \AA\ break and G band absorption line.  
Fortunately, the \ion{Ca}{2} H\&K feature fell 
150\AA\ redward of the [\ion{O}{1}] $\lambda$5577 skyline, which provided 
points for wavelength calibration on either side of the feature.

To determine the galaxy redshift, we simultaneously fit for the the 
continuum plus two overlapping Gaussian profiles to the \ion{Ca}{2} H\&K 
absorption feature and bracketing continuum using an appropriately fixed 
rest-wavelength separation.  This does a good job of modeling the absorption 
feature (reduced chi-squared of 0.7) and yields a best-fit redshift of 
$z=0.4546$.  The uncertainty in the fit is of the same order as the error in 
the wavelength calibration, $\sim$ 1 \AA, translating to a redshift 
uncertainty of 0.0002.

Figure~2 also shows the extracted spectra for galaxies G12 and G13, the
two closest galaxies from the lens that we obtained spectra for
($\Delta\theta$=8\farcs9 and $\Delta\theta$=31\farcs4 from the lens, 
respectively).  Both show [\ion{O}{2}]$\lambda$3727 emission at 
$\sim$5300\AA, and G12 also shows 
weak H$\beta$ and [\ion{O}{3}]$\lambda\lambda$4959,5007 emission lines at 
$\sim$7000\AA.  Gaussian fits to the emission lines give redshifts of
$z=0.4191\pm0.0002$ and $z=0.4189$ for G12 and G13, respectively.

Table~1 lists the spectroscopic redshifts obtained for all 18 galaxies
observed with LDSS2, and Figure~3 shows the redshift histogram.  There is no 
obvious group or cluster along the line of sight from the limited redshift 
sample, although the coincident G12,G13 redshifts at $z=0.419$ suggests 
some overdensity at that redshift.  The redshift difference between the 
G12,G13 pair and the lensing galaxy is 7,500 km s$^{-1}$ in the G12,G13
restframe, an order of magnitude larger than expected for cluster association 
and even more so for a group, so the galaxies are not physically associated.  
The closest galaxy in redshift space to the lens is G11 
($z=0.4579$) at a projected separation of $\Delta\theta$=74\farcs6 
(proper distance of 420 kpc), which is too far to significantly affect 
the lensing potential.

\section{{\it HST}/ACS Observations}
\label{hstacs}

HE~0435$-$1223 was observed with the Advanced Camera for Surveys/Wide Field 
Camera (ACS/WFC; Ford et al. 1998) onboard the {\it Hubble Space Telescope} on 
2003 August 18 as part of the CASTLES imaging program of gravitationally lensed
quasars (principal investigator, C. Kochanek; PID 9744).  Five images each 
through the F555W and F814W filters were obtained (hereafter $V$ and $I$) with
respective total integration times of approximately 34 and 24 minutes.  The 
data were reduced using the IRAF/CALACS\footnote{IRAF is distributed by the 
National Optical Astronomy Observatories, which are operated by the 
Association of Universities for Research in Astronomy, Inc., 
under cooperative agreement with the National 
Science Foundation.} package as part of the ``on-the-fly'' reprocessing at the 
time of download.  Subsequent cosmic-ray rejection, geometric correction, and 
image combinations were performed using the standard PyRAF programs available 
for ACS data reduction.

Figure~4 shows the geometrically-corrected $I$-band image after combining with 
PyDrizzle.  The environment around the lens appears to contain a spiral-rich 
group of galaxies.  There are 5 prominent galaxies within 10\arcsec\ (outer 
dashed circle), the closest of which (G22) is within 5\arcsec\ and corresponds 
to the SW companion galaxy originally noted by Wisotzki et al. (2002).  Its 
image morphology is a barred spiral with arms that appear to thread back onto
the central bulge.  Galaxies G12 and G13 from \S\ref{ldss2} are also 
labeled.  Both show a bulge+disk image morphology, face-on in the case of G12 
and edge-on in the case of G13, with several bright knots visible distinct from
the central nucleus in both galaxies.  The image morphologies are consistent 
with ongoing star formation as implied by the 
[\ion{O}{2}]$\lambda$3727 \AA\ emission-lines observed in the respective 
spectra.  Galaxies G18, G19, G23, and G25 also suggest either face-on or 
edge-on spirals, while G20, G21, and G24 appear to be ellipticals.

\begin{deluxetable*}{ccccccc}
\tabletypesize{\scriptsize}
\tablecaption{{\it HST}/ACS Astrometry and Photometry for HE~0435$-$1223
\label{TABLE2}}
\tablewidth{0pt}
\tablehead{  
\colhead {Object}&
\colhead {$\Delta$ R.A. (\arcsec)}&
\colhead {$\Delta$ Dec. (\arcsec)}&
\colhead {$V$/$V_A$}&
\colhead {$I$/$I_A$}&
\colhead {$V$-band}&
\colhead {$I$-band}
}
\startdata
A\dotfill &   $\equiv$0            &   $\equiv$0            &   $\equiv$1       &   $\equiv$1       & 18.545 $\pm$ 0.001 & 19.106 $\pm$ 0.001 \nl
B\dotfill &  $-$1.4772 $\pm$ 0.002 &  +0.5532 $\pm$ 0.002   & 0.607 $\pm$ 0.016 & 0.621 $\pm$ 0.003 & 19.082 $\pm$ 0.001 & 19.623 $\pm$ 0.001 \nl
C\dotfill &  $-$2.4687 $\pm$ 0.002 &  $-$0.6033 $\pm$ 0.002 & 0.579 $\pm$ 0.012 & 0.617 $\pm$ 0.003 & 19.149 $\pm$ 0.001 & 19.631 $\pm$ 0.001 \nl
D\dotfill &  $-$0.9377 $\pm$ 0.002 &  $-$1.6147 $\pm$ 0.002 & 0.557 $\pm$ 0.010 & 0.516 $\pm$ 0.003 & 19.196 $\pm$ 0.001 & 19.824 $\pm$ 0.001 \nl
G\dotfill &  $-$1.1687 $\pm$ 0.002 &  $-$0.5723 $\pm$ 0.002 &   $\cdots$        &   $\cdots$        &  20.80 $\pm$ 0.10  &  19.83 $\pm$ 0.08  \nl
\enddata
\tablecomments{Relative image positions and system flux ratios are from PSF 
fitting as described in the text.  Quasar apparent magnitudes are from 0\farcs5 
aperture photometry plus a nominal 0.1 mag aperture correction.  Galaxy magnitudes 
have been obtained using the best-fit $I$-band de Vaucouleurs profile, with error 
bars corresponding to 10\% uncertainties in $r_e$.  All magnitudes have been 
placed onto the STMAG system.
}
\end{deluxetable*}

The $V$- and $I$-band closeup images of HE~0435$-$1223 are shown in Figure~5a 
and 5b.  The image configuration is consistent with previous ground-based 
images of the system, although the {\it HST} data do reveal multiple partial 
arcs tracing the system's Einstein radius in both the $V$- and $I$-band images.
One can also see lensed knots at two different radii embedded in the 
arc emission and which likely arises from lensed structure of the quasar host 
galaxy.  To obtain the relative image positions, we modeled the light 
distribution using PSFs generated with the TinyTim v6.1a software of Krist 
\& Hook (2003) and a circularly symmetric de Vaucouleurs profile (with 
effective radius of 1\farcs2; see below) convolved with the {\it HST} PSF for 
the lensing galaxy.  The PSFs were generated taking into account the lens 
position on the WFC and mapped onto a 4x4 oversampled grid to assist with 
subpixel shifts.  The relative positions and fluxes of the five-component model
were then simultaneously solved for using a Powell (Press et al. 1992) 
minimization routine.  The solutions were obtained in the two filters using 
both the individual (geometrically distorted) `flt' images and the undistorted 
drizzled frames, and we found that the relative quasar positions agreed to 
within 0\farcs002 using the different datasets.  The galaxy center rms was 
0\farcs005 (0\farcs002) along the cardinal directions for the $V$($I$)-band 
solutions, roughly an order of magnitude improvement over the earlier
position obtained from ground-based observations.  

The averaged relative offsets obtained from the `flt' $I$-band solutions are 
reported in Table~2.  To our initial surprise, the relative quasar positions 
differed by up to 6$\sigma$ when compared to the Magellan offsets given by 
Wisotzki et al. (2002), even though both datasets quote similar astrometric 
precision.  The differences, however, could be described by a 0.6\% 
scale discrepancy between the two datasets, in the sense that the Wisotzki 
et al. (2002) measurements correspond to a larger angular size than the 
{\it HST}/ACS measurements.  After correcting for the scale difference, the 
Magellan and {\it HST} astrometry agree to better than 1$\sigma$.  

The reconstructed $I$-band image without the four quasars (Figure~5c) allows 
us to investigate the detailed morphological properties of the lensing galaxy.
We modeled the galaxy's $I$-band light distribution of the PSF-subtracted image
using {\tt GALFIT} (Peng et al. 2002), taking care to mask the ring of pixels 
tracing the quasar residuals and arc emission during the fit.  The best-fit 
elliptical $r^{1/4}$ law gave an axis ratio of 0.83, a position angle of 
$-$7\fdg4 East of North, and an effective radius $r_e$ of 1\farcs20 (6.6 kpc 
at the lens redshift). Figure~5d shows the residuals after subtracting the 
best-fit profile.  

\begin{deluxetable*}{ccccc}
\tablecaption{{\it HST}/ACS Astrometry and Photometry for HE~0435$-$1223 Field
\label{TABLE3}}
\tablewidth{0pt}
\tablehead{  
\colhead {Object}&
\colhead {$\Delta$ R.A. (\arcsec)}&
\colhead {$\Delta$ Dec. (\arcsec)}&
\colhead {$V$-band}&
\colhead {$I$-band}
}
\startdata
G12\dotfill  &  \0$-$8.96  &  \p\03.66   &  21.182 $\pm$ 0.018  &  21.158 $\pm$ 0.009  \nl
G13\dotfill  &   $-$30.18  &   \p11.42   &  22.486 $\pm$ 0.047  &  21.980 $\pm$ 0.015  \nl
G18\dotfill  &   $-$20.37  &   \p29.56   &  22.147 $\pm$ 0.043  &  21.829 $\pm$ 0.012  \nl
G19\dotfill  &   $-$27.69  &   \p18.01   &  21.517 $\pm$ 0.019  &  21.230 $\pm$ 0.009  \nl
G20\dotfill  &  \0$-$6.74  &   \p13.64   &  23.946 $\pm$ 0.155  &  22.776 $\pm$ 0.033  \nl
G21\dotfill  &  \0$-$7.62  &  \p\06.61   &  22.564 $\pm$ 0.056  &  21.819 $\pm$ 0.013  \nl
G22\dotfill  &  \0$-$3.75  & \0$-$4.21   &  22.253 $\pm$ 0.042  &  21.260 $\pm$ 0.010  \nl
G23\dotfill  &  \0$-$5.39  & \0$-$9.46   &  24.409 $\pm$ 0.204  &  22.850 $\pm$ 0.035  \nl
G24\dotfill  &   \p\01.01  & \0$-$8.39   &  23.946 $\pm$ 0.155  &  22.501 $\pm$ 0.024  \nl
G25\dotfill  &   $-$18.36  &  $-$18.25   &  22.134 $\pm$ 0.025  &  21.910 $\pm$ 0.014  \nl
 & & & & \nl
S1\dotfill  &  \p\09.48  & \0$-$7.16   &  21.415 $\pm$ 0.004  &  20.258 $\pm$ 0.002  \nl
S2\dotfill  &   \p11.59  &  $-$51.29   &  20.183 $\pm$ 0.002  &  19.250 $\pm$ 0.001  \nl
S3\dotfill  &  $-$41.98  & \p67.73     &  20.152 $\pm$ 0.002  &  20.575 $\pm$ 0.002  \nl
\enddata
\tablecomments{Galaxy magnitudes were obtained using a 2\farcs0 aperture radius, with error bars 
corresponding to Poisson noise inside the aperture.  Stellar magnitudes are from 0\farcs5 aperture 
photometry plus a nominal 0.1 mag aperture correction.  All magnitudes have been placed onto the 
STMAG system.
}
\end{deluxetable*}

The effective surface brightness of the galaxy $\mu_e$, defined as
the average flux inside the effective radius, is 22.22 mag in $I$-band.  
To get a handle on the $V$-band effective surface brightness, we applied 
the same $I$-band effective radius and normalized the de Vaucouleurs profile 
using the PSF-subtracted $V$-band image.  This gave $\mu_e$ = 23.19 mag
in $V$-band.  The total lens galaxy magnitude in the two filters then follows 
from $m_{tot} = \mu_e - 5.0\log r_e - 2.5\log 2\pi$ and are reported in 
Table~2.

In Table~3, we list apparent magnitudes and relative positions for the 10 
prominent galaxies labeled in Figure~4 and for three nearby reference 
stars.  Galaxy aperture magnitudes were computed using a 2\farcs0 radius with 
error bars estimated from the Poisson noise inside each aperture.  Stellar
magnitudes were computed using a 0\farcs5 aperture radius and a nominal
0.1 mag aperture correction.

\section{Lens Models and Time Delay Predictions}
\label{models}

Predicting differential time delays between the quasar images requires 
an accurate mass model of the lensing galaxy and its environment.  Wisotzki et 
al. (2002) found that a simple singular isothermal sphere (SIS) embedded 
in an external shear did a good job at modeling the quasar positions 
(rms of 1.3 mas), but failed to account for the quasar flux ratios.  
When compared to their $i$-band fluxes, the SIS+shear model (which only used 
position constraints) underpredicted component A's flux by 0.35 mag.
Overall, it predicted components B+C to be almost a full magnitude (0.86 mag)
brighter relative to components A+D than observed.

Such discrepancies are common when modeling strong lenses, and it is 
generally accepted that the image fluxes are perturbed either by substructure 
along the line of sight (Dalal \& Kochanek 2002) or by microlensing from stars
in the lensing galaxy (Schechter \& Wambsganss 2002).  These two processes
are in principle distinguishable from each other since they yield different
predictions for extended sources.  The flux ratios ought not to show a 
strong dependence on source size if significant substructure is present, but 
will show a source size effect for microlensing-induced perturbations.  This 
is because larger source sizes will average over small-scale structure present
in the microlensing caustic pattern, driving the system flux ratios toward the
macromodel values.  In practice, since the quasar broad-line region is
several orders of magnitude larger than the continuum emitting region, 
then the emission-line fluxes ought to be much less sensitive to microlensing 
than the broad-band fluxes.

The \ion{C}{4} and \ion{C}{3}] emission-line fluxes for HE~0435$-$1223
were measured by Wisotzki et al. (2003) using the Potsdam Multi-Aperture 
Spectrophotometer (PMAS) on the Calar-Alto 3.5~m telescope.  They indeed 
found that the emission-line flux ratios were in better agreement 
with model predictions than were the broadband values, but the match
was still far from acceptable: component A was underpredicted by 0.26 mag and
the combined B+C flux was still 0.57 mag brighter with respect to components 
A+D than observed.  Attempts to force the SIS+shear model to reproduce the 
emission-line flux ratios could be achieved only with significant ($\sim 
50\sigma$) deviations between the observed and model image positions.  This 
difficulty could be a consequence of the simple isothermal model used for
the lensing potential and a more realistic model, perhaps taking the group 
environment into account, might be more successful. 

In this section, we explore several mass models for the system in an
effort to reproduce the {\it HST}/ACS astrometry and the Wisotzki 
et al. (2003) emission-line photometry.  Our basic mass model is a
singular isothermal ellipsoid (SIE; Kassiola \& Kovner (1993), Kormann et al. (1994)) 
with convergence $\kappa$ given
by 
\begin{equation}
\kappa = \frac{\Sigma}{\Sigma_{c}} = \frac{b}{2r}\left(\frac{1}{1 + \epsilon \cos 2\left(\theta-\theta_e\right)}\right)^{1/2},
\end{equation}
where $\Sigma$ is the projected surface mass density, $\Sigma_{c} 
= (c^2/2\pi G)(D_d/D_lD_{ls})$ is the critical surface mass density, $\epsilon$
is the ellipticity parameter related to the axis ratio $q$ by $q^2 = (1-\epsilon)/(1+\epsilon)$,
and the major axis is orientated along $\theta_e$.  
The mass parameter $b$ is given by
\begin{equation}
b = 4\pi \left(\frac{\sigma}{c}\right)^2\frac{D_{ls}}{D_s},
\end{equation}
where $\sigma$ is the line-of-sight velocity dispersion of the dark matter 
halo.  Distances $D_l$, $D_s$, and $D_{ls}$ are angular diameter distances
to the lens, the source, and from the lens to source, respectively.  For the 
SIS model ($q=1$), $b$ is also the 
system's Einstein radius.  The two-dimensional effective lensing potential 
$\phi$ can then be found from $\nabla^2\phi = 2\kappa$, and is simply
$\phi = br$ for the spherically symmetric case (e.g., Narayan \& Bartelmann 1999).  
We also consider environmental effects by adding a shear term $\phi_{\gamma}$ 
of the form
\begin{equation}
\phi_{\gamma} = - \gamma\frac{r^2}{2} \cos 2\left(\theta-\theta_{\gamma}\right),
\end{equation}
where $\gamma$ is the shear strength and the sign convention is for 
$\theta_{\gamma}$ to point toward (or away) from the mass responsible for the 
shear.  

We consider three models, defined as follows:
\begin{eqnarray}
\phi_{ISx} &=& \phi_{SIS} + \phi_{\gamma} \\
\phi_{IEx} &=& \phi_{SIE} + \phi_{\gamma} \\
\phi_{IEISx} &=& \phi_{SIE} + \phi_{SIS,G22} + \phi_{\gamma}.
\end{eqnarray}
Model ISx is a singular isothermal sphere for the lensing galaxy plus 
external shear.  It is the same model used by Wisotzki et al. (2002).  
Model IEx adds a second shear axis by using an elliptical isothermal halo 
plus external shear.  Model IEISx explicitly takes the galaxy environment 
into account.  It adds a third shear axis by fixing a second isothermal halo 
at the position of the closest neighboring galaxy G22.  

\begin{deluxetable*}{lcccccccc}
\tablecaption{Lens Model Results for HE~0435$-$1223
\label{TABLE4}}
\tablewidth{0pt}
\tablehead{  
\colhead{} & \multicolumn{2}{c}{ISx} & \colhead{} & \multicolumn{2}{c}{IEx} & \colhead{} & \multicolumn{2}{c}{IEISx} \\
\cline{2-3} \cline{5-6} \cline{8-9}  \\
\colhead{Parameter} & \colhead{10\%} & \colhead{1\%} & \colhead{} & \colhead{10\%} & \colhead{1\%} & \colhead{} & \colhead{10\%} & \colhead{1\%} \\
}
\startdata
                              &               &               & &               &               & &               &              \\
$\chi^2$/dof\dotfill          & 21.3/6        & 816/6         & & 9.7/4         & 134/4         & & 5.8/3         & 87.3/3       \\
$\chi^2_{pos}$\dotfill        & \07.1         & 28.9          & & \00.0         & 32.3          & & \00.0         & 13.3         \\
$\chi^2_{flux}$\dotfill       & \08.3         & 775\0\0       & & \08.6         & 43.3          & & \05.5         & 15.9         \\
$\chi^2_{gal}$\dotfill        & \05.9         & 11.3          & & \01.1         & 59.0          & & \00.2         & 58.4         \\
$\theta_{rms}$ (mas)\dotfill  & \02.7         & \05.4         & & \00.2         & \05.7         & & \00.2         & \03.7        \\
                              &               &               & &               &               & &               &              \\
$\Delta m_A$ (mag)\dotfill    & \m0.28        & \m0.28        & &  \m0.25       &  \m0.05       & &  \m0.20       &  \m0.03      \\
$\Delta m_B$ (mag)\dotfill    & $-$0.10       & $-$0.09       & &  $-$0.07      &  $-$0.04      & &  \m0.01       &  $-$0.01     \\
$\Delta m_C$ (mag)\dotfill    & $-$0.13       & $-$0.13       & &  $-$0.16      &  $-$0.03      & &  $-$0.16      &  \m0.00      \\
$\Delta m_D$ (mag)\dotfill    & \m0.08        & \m0.07        & &  \m0.12       &  \m0.03       & &  \m0.05       &  $-$0.02     \\
                              &               &               & &               &               & &               &              \\
$b\h$ (arcsec)\dotfill        & \p1.20        & \p1.20        & & \p1.20        & \p1.17        & & \p1.17        & \p1.17       \\
$b'$ (arcsec)\dotfill         & \p$\cdots$    & \p$\cdots$    & & \p$\cdots$    & \p$\cdots$    & & \p0.18        & \p0.30       \\
$\gamma$\dotfill              & \p0.08        & \p0.08        & & \p0.01        & \p0.14        & & \p0.04        & \p0.14       \\
$e$\dotfill                   &  \p$\cdots$   & \p$\cdots$    & & \p0.13        & \p0.30        & & \p0.22        & \p0.29       \\
$\theta_{\gamma}$\dotfill     & $-$13.8       & $-$13.8       & & $-$15.9       & $-$30.0       & & $-$28.0       & $-$27.3      \\
$\theta_e$\dotfill            &  \p$\cdots$   & \p$\cdots$    & & $-$11.5       & \p60.0        & & $-$14.1       & \p60.2       \\
                              &               &               & &               &               & &               &              \\
$\tau_{B-A}$ (days)\dotfill   &  \05.48       &  \05.40       & &  \07.18       &  \p1.88       & &  \08.72       & \p1.87       \\
$\tau_{C-A}$ (days)\dotfill   &  \00.61       &  \00.44       & &  \00.85       & $-$0.14       & &  \00.70       & $-$0.18      \\
$\tau_{D-A}$ (days)\dotfill   & 12.21         & 11.89         & & 15.80         &  \p4.25       & & 17.54         & \p4.07       \\

\enddata
\tablecomments{Model results for the isothermal sphere plus external shear (ISx),
isothermal ellipsoid plus external shear (IEx), and isothermal ellipsoid 
plus external shear and isothermal sphere centered on G22 (IEISx).  Models are minimized
using 10\% and 1\% emission-line flux errors.  $\theta_{rms}$ gives the rms between predicted
and observed quasar images.  Mass parameters $b$ and $b'$ are for the lens and G22 galaxies, 
respectively.  Magnitude differences are in the sense predicted minus observed.}
\end{deluxetable*}

For constraints, we use the {\it HST}/ACS positions listed in Table~2 with 2 
mas error circles for the quasar positions and a 5 mas error circle for the 
galaxy position.  The lens galaxy position is allowed to vary, but its 
position is constrained using the above error circle.  The averaged \ion{C}{4}
and \ion{C}{3}] emission-line fluxes can be computed from Table~2 of Wisotzki 
et al. (2003) and give B/A, C/A, and D/A ratios of 0.76, 0.71, and 
0.54, respectively.  Wisotzki et al. (2003) quote formal errors for the 
emission-line fluxes on the order of $1\%$.  While we consider models with 
1\% flux errors, we also consider models with 10\% flux errors to account
for unmodeled effects such as intrinsic quasar variability.  We also assume 
identical redshifts for all deflectors,
even though the LDSS2 spectroscopy from \S\ref{ldss2} has already showed that 
this is not strictly the case ($\Delta z = 0.036$ for the lensing galaxy and 
G12,G13).  Throughout this section, we adopt an ($\Omega_m,\Omega_{\Lambda}$) 
= (0.3, 0.7) cosmology, set the Hubble constant to 72 km s$^{-1}$ 
Mpc$^{-1}$, and quote all angular positions as degrees East of North.  
All models are minimized in the image-plane using the {\tt gravlens} 
software of Keeton et al. (2001).  Model results are summarized in Table~4.

\subsection{Single Halo: Models ISx and IEx}

Wisotzki et al. (2003) found that an isothermal halo plus external shear
cannot simultaneously fit the Magellan positions and emission-line flux 
ratios.  We find a similar result using the {\it HST}/ACS positions as 
constraints.  Using 10\% flux errors, the ISx model does well to
reproduce the image positions ($\theta_{rms} = 2.7$ mas), but 
predicts components B+C to be 0.59 mag brighter with respect to components A+D
than observed.  In particular, component A's model flux is 0.28 mag fainter 
than observed.  This is essentially the same result found by Wisotzki et al. 
(2003) using just the Magellan image positions as constraints.  Tightening the
flux errors to 1\% does not yield an improvement: $\theta_{rms}$ worsens to 
5.4 mas and the image magnifications are mostly unchanged (see Table~4).

The IEx model allows for two shear axes, external shear plus galaxy 
ellipticity.  One has to be careful when solving for the shear and 
ellipticity terms since an approximate degeneracy exists between the two 
effects (Keeton et al. 1997).  To sift through the degeneracy, we solved for 
($b$, $\gamma$, $e$) over a grid of initial conditions for 
($\theta_{\gamma}, \theta_e$) and focused on the region of parameter space 
that gave the lowest $\chi^2$ values.  The best-fit model using 10\% flux 
errors fit the quasar positions essentially exactly ($\theta_{rms} = 0.2$ 
mas), but again had trouble predicting the flux ratios, with the combined B+C 
flux 0.61 mag brighter with respect to A+D than observed.  The problem again 
is chiefly component A, which is predicted to be 0.25 too faint.  The galaxy 
parameters do agree with the observed characteristics: the 
model ellipticity $e$ (defined as 1 minus the axis ratio) was 0.13 and pointed 
along $-$11\fdg5, which is close to the measured values of 0.17 and $-$7\fdg4 
for the light distribution.  

Tightening the IEx flux errors to 1\% improved the flux predicitions --- 
within 0.05 mag for all four components --- but the galaxy parameters became 
unreasonable: the ellipticity was 0.29, almost twice as large than observed, 
and pointed nearly perpendicular ($\Delta\theta=70\degr$) from the observed 
orientation.  Moreover, the galaxy position shifted by 0\farcs038 
(7-8$\sigma$) from its measured center.  Overall the model can be rejected
since it predicts halo properties in strong disagreement with the observed 
galaxy properties.  Thus neither of the single halo models
are consistent with both the observed galaxy properties and the system's
emission-line flux ratios.  

\subsection{Multiple Halos: Model IEISx}

The shear direction for the ISx and IEx models pointed roughly at $-15\degr$.
Naively one would expect it to point toward the closest neighboring galaxy 
G22 at 216\degr\ (or equivalently 36\degr).  The same difference was found 
Wisotzki et al. (2002) and led them to speculate that the neighboring galaxy 
might simply be a chance projection.  Another possibility is that G22 does 
have a significant effect on the lensing potential, but that a second 
perturber NW of the lens dominates the shear contribution.  Figure~4 shows 
that this is plausible.  Galaxies G12, G20, and G21 (as well as G18, G19 and 
G13 farther out) range from position angles of $-$20\degr\ to $-$70\degr\ and 
their aggregate mass or associated halo may dominate the shear measurement.  

Model IEISx takes both shear directions into account by using a halo centered 
on G22 for the SW perturber plus an external shear for the NW perturber.  
Using 10\% flux errors, the model does just as well as the IEx model in 
predicting the image positions ($\theta_{rms}$ = 0.2 mas) and also slightly 
improves the flux predictions.  This time the B+C flux was only 0.40 mag 
brighter with respect to A+D than observed, component A was too faint by 
0.20 mag, and component C was too bright by 0.16 mag.  The galaxy 
ellipticity was also slightly larger than observed, 0.22 compared to 0.17, 
but did point within 7\degr\ of the measured orientation.  
The shear strength for the 10\% model was 0.04 and pointed at $-$28\degr, or 
about 13\degr\ closer to the G20-G21-G12 group than found in the 
single halo IEx model.  We again reject the 1\% model since it gives 
unreasonable galaxy parameters (see Table~4) despite the improved image fluxes.

\section{Discussion}
\label{discuss}

\begin{figure}[t]
\plotone{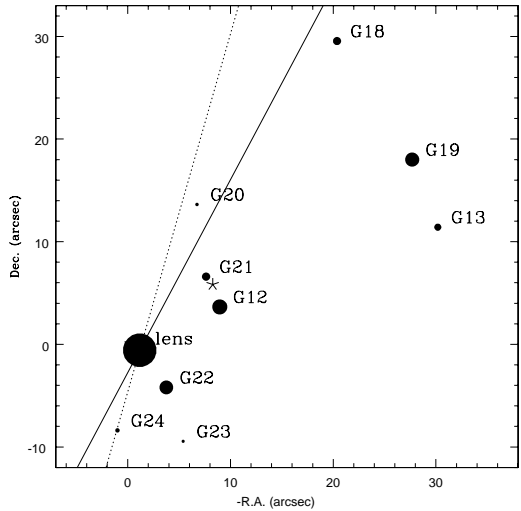}
\caption{Diagram showing relative galaxy positions (filled circles).
Dot sizes are proportional to the respective galaxy's $I$-band flux within
a 2\arcsec\ aperture.  The diagonal lines show the shear directions for the
IEx (dotted) and IEISx (solid) lens models.  The star marks the flux-weighted
center of the G20-G21-G12 group of galaxies.}
\label{fig:shear}
\end{figure}

The improved flux predictions and closer alignment between the shear direction
and the NW group are arguments for favoring the multiple halo 
model, but the shear still does not point to any obvious perturber.  It points 
between the faint G20 and G21 galaxies as seen from the lens.  This is 
about +20\degr\ from the G20-G21-G12 flux-weighted center and about +30\degr\ 
from G12, the brightest of all neighboring galaxies within 40\arcsec\ of the 
lens (see Figure~6).  We {\it know} that G12 is at essentially the same 
redshift as the lensing galaxy, so if the only significant perturbers were 
G12 to the NW and G22 to the SW, then we would expect the IEISx shear to 
point directly toward G12.  It does not, which suggests that if the flux 
ratios are to be explained solely by the macromodel, then the lensing 
potential must be even more complicated than the double halo model considered 
here.

Alternatively, the macromodel could be essentially correct and a separate 
effect could be perturbing the flux ratios.  In general, there are four 
mechanisms that could alter the image fluxes: microlensing, quasar 
variability, differential dust extinction and substructure lensing.  
Presumably microlensing is not an issue since we are working with the 
emission-line fluxes.  Quasar variability is also not a factor since the 
A$-$C time delay is less than a day in all models, much too short compared to 
quasar variability timescales (typically weeks to months).

Wisotzki et al. (2002) already commented that extinction is probably not very 
important for this system since the lensing galaxy is an elliptical and the 
Magellan colors of the four quasars agreed to $\lesssim$0.05 mag.  
In general, Falco et al. (1999) found median differential extinctions among 
a subsample of optically-selected lenses of $\Delta E(B-V)=0.04$ mag.  Only 
10\% of their sightlines through elliptical galaxies have $0.15<\Delta 
E(B-V)<0.25$, so the $\sim$0.2 mag of extinction needed to reconcile 
component C's relative faintness is unlikely but not impossible.   However, 
the PMAS resolved spectral observations of the four quasars obtained by 
Wisotzki et al. (2003) show the quasar continuum slopes to be virtually 
identical (L. Wisotzki, private communication).  This would be a strong 
coincidence if any significant dust extinction were present, and of course 
rules out significant differential extinction.

The remaining explanation is substructure lensing.  Millilensing by low-mass 
satellites (typically modeled as NFW profiles or truncated 
isothermal halos) in the lens galaxy can either brighten or dim image fluxes 
by several tenths of a magnitude.  The effect has been modeled
both for individual systems (Metcalf \& Zhao 2002) and in 
a statistical sense (Dalal \& Kochanek 2002), and typically
requires the subclumps to comprise a few percent by mass of the 
primary galaxy's dark-matter halo.  Although we do not consider substructure 
models here, such an effect remains the most plausible explanation of the 
HE~0435$-$1223 emission-line fluxes given the present data.  

We do note that substructure lensing by strictly isothermal halos, either 
embedded in the lensing galaxy or along the line of sight, cannot account for 
the observed flux ratios.  Isothermal mass clumps always brighten 
positive-parity images (Keeton 2003), but components A and C, both 
positive-parity images, are respectively brighter and fainter than predicted. 
Of course the unlensed quasar flux is a free parameter, so one could
dim the source such that both A and C became brighter than predicted, but
then images B and D (both negative-parity images) would be $\sim$0.2 mag
brighter than predicted as well.  Millilensing (and microlensing as well) 
tends to dim saddle-point images (Keeton 2003, Schechter \& Wambsganss 2002),
so the situation using isothermal halos becomes somewhat contrived.  

We can predict the differential time delays for the system using the measured
lens redshift of $z=0.4546\pm0.0002$ from \S\ref{ldss2}.  The A$-$D delay is 
the longest delay and therefore the most interesting to measure.  The ISx, 
IEx, and IEISx models predict A$-$D delays of 12.21, 15.80, and 17.54 days, 
respectively.  The elliptical halo (IEx) is formally preferred over the 
spherical halo (ISx) with $\chi^2/dof$ of 3.6 compared to 2.4, where the 
improvement comes wholly from the quasar and galaxy positions.  The double
halo (IEISx) is preferred over the single halo (IEx) with $\chi^2/dof$ of 
2.4 compared to 1.9, where the improvement comes mostly from the images 
fluxes.  The last two models yield 10\% differences in the A$-$D time delay, 
which gives an estimate of the error that one can expect from uncertainties 
in the local galaxy environment.

There are several ways to improve the time delay prediction.  Redshift 
information for the remaining 8 of 10 neighboring galaxies, especially for G22,
would significantly improve the model confidence.  In the least, it would 
point to which galaxies might share a common dark matter halo and thus guide 
the choice of parametric models.  The {\it HST}/ACS images already show a 
partial Einstein ring for the system in $I$-band, and one would suspect
a complete Einstein ring to be visible in the infrared.  The extra 
constraints offered by modeling the ring may be able to distinguish between 
the single and double halo models without recourse to the image flux ratios.  
For example, it would be interesting to see if the shear direction for a 
ring-constrained IEISx model points more in line with the center of the NW 
group than found here.

\acknowledgments

This work is based on observations made with the NASA/ESA {\it Hubble Space 
Telescope}, obtained at the Space Telescope Science Institute, which is 
operated by AURA, Inc., under NASA contract NAS5-26555.  This research 
is supported by HST grants G0-9375 and GO-9744.  PLS acknowledges support 
from the US NSF under AST02-06010.


\begin{thebibliography}{}

\bibitem[Allington-Smith et al. (1994)]{alli94} Allington-Smith, J., Breare, M., Ellis, R., Gellatly, D., Glazebrook, K., Jorden, P., Maclean, J., Oates, P., Shaw, G., Tanvir, N., Taylor, K., Taylor, P., Webster, J., Worswick, S. 1994, \pasp, 106, 983

\bibitem[Dalal \& Kochanek (2002)]{koch02} Dalal, N. \& Kochanek, C. S. 2002, \apj, 572, 25

\bibitem[Falco et al. (1999)]{falc99} Falco, E. E., Impey, C. D., Kochanek, C. S., Leh\'ar, J., McLeod, B. A., Rix, H.-W., Keeton, C. R., Mu\"noz, J. A. \& Peng, C. Y. 1999, \apj, 523, 617

\bibitem[Ford et al. (1998)]{ford98} Ford, H. C., Bartko, F., Bely, P. Y., Broadhurst, T., Burrows, C. J., Cheng, E. S., Clampin, M., Crocker, J. H., Feldman, P. D., Golimowski, D. A., Hartig, G. F., Illingworth, G., Kimble, R. A., Lesser, M. P., Miley, G., Neff, S. G., Postman, M., Sparks, W. B., Tsvetanov, Z., White, R. L., Sullivan, P., Krebs, C. A., Leviton, D. B., La Jeunesse, T., Burmester, W., Fike, S., Johnson, R., Slusher, R. B., Volmer, P., Woodruff, R. A. 1998, Proc. SPIE, 3356, 234-248

\bibitem[Gorenstein, Shapiro \& Falco (1988)]{gore88} Gorenstein, M. V., Shapiro, I. I. \& Falco, E. E. 1988, \apj, 327, 693

\bibitem[Holder \& Schechter (2003)]{hold03} Holder, G. \& Schechter, P. L. 2003, \apj, 2003, 688

\bibitem[Huchra et al. (1985)]{huch85} Huchra, J., Gorenstein, M., Kent, S., Shapiro, I., Smith, G., Horine, E., Perley, R. 1985, \aj, 90, 691

\bibitem[Kassiola \& Kovner (1993)]{kass93} Kassiola, A. \& Kovner, I. 1993, \apj, 417, 450

\bibitem[Keeton et al. (1997)]{keet97} Keeton, C. R., Kochanek, C. S. \& Seljak, U. 1997, \apj, 482, 604

\bibitem[Keeton (2001)]{keet01} Keeton, C. R. 2001 (astro-ph 0102340)

\bibitem[Keeton (2003)]{keet03} Keeton, C. R. 2003, \apj, 584, 664

\bibitem[Keeton \& Zabludoff (2004)]{keet04} Keeton, C. R. \& Zabludoff, A. I. 2004, \apj, 612, 660

\bibitem[Kormann et al. (1994)]{korm94} Kormann, R., Schneider, P. \& Bartelmann, M. 1994, \aap, 284, 285

\bibitem[Krist \& Hook (2003)]{kris03} Krist, J. E. \& Hook, R. N. 2003, The Tiny Time User's Guide, Version 6.1a (Baltimore: STScI)

\bibitem[Metcalf \& Zhao (2002)]{metc02} Metcalf, R. B. \& Zhao, H. 2002, \apj, 567, L5

\bibitem[Narayan \& Bartelmann (1999)]{nara94} Narayan, R. \& Bartelmann, M. 1999, in Formation of Structure in the Universe, ed. A. Dekel \& J. Ostriker (Cambridge: Cambridge Univ. Press)

\bibitem[Osterbrock et al. (1996)]{oste96} Osterbrock, D. E., Fulbright, J. P., Martel, A. R., Keane, M. J., Trager, S. C., Basri, G. 1996, \pasp, 108, 277

\bibitem[Peng et al. (2002)]{peng02} Peng, C. Y., Ho, L. C., Impey, C. D. \& Rix, H.-W. 2002, \aj, 124, 266

\bibitem[Press et al. (1992)]{pres92} Press, W. H., Teukolsky, S. A., Vetterling, W. T. \& Flannery, B. P. 1992, Numerical Recipes in C (Cambridge: Cambridge Univ. Press)

\bibitem[Refsdal (1964)]{refs64} Refsdal, S.  1964, \mnras, 128, 307

\bibitem[Rusin et al. (2003)]{rusi03} Rusin, D., Kochanek, C. S., Falco, E. E., Keeton, C. R., McLeod, B. A., Impey, C. D., Leh\'ar, J., Mu\~noz, J. A., Peng, C. Y., Rix, H.-W.  2003, \apj, 587, 143

\bibitem[Saha (2000)]{saha00} Saha, P. 2000, \aj, 120, 1654

\bibitem[Schechter \& Wambsganss (2002)]{sche02} Schechter, P. L. \& Wambsganss, J. 2002, \apj, 580, 685

\bibitem[Wisotzki et al. (2000)]{wiso00} Wisotzki, L., Christlieb, N., Bade, N., Beckmann, V., K\"ohler, T., Vanelle, C., Reimers, D. 2000, \aap, 358, 77

\bibitem[Wisotzki et al. (2002)]{wiso02} Wisotzki, L., Schechter, P. L., Bradt, H. V., Heinm\"uller, J., Reimers, D. 2002, \aap, 395, 17

\bibitem[Wisotzki et al. (2003)]{wiso03} Wisotzki, L., Becker, T., Christensen, L., Helms, A., Jahnke, K., Kelz, A., Roth, M. M., Sanchez, S. F. 2003, \aap, 408, 455

\end{thebibliography}
\end{document}